\documentclass{aagiallo} 
\newcommand{\be} {\begin{equation}} 
\newcommand{\ee} {\end{equation}} 
\newcommand{\src}{GS\,1843+009} 
\newcommand{\srchri}{1BMW\,J184536.8+005148} 
\newcommand{\R}{{\em ROSAT}}

\newcommand{\bc}{\begin{center}} 
\newcommand{\ec}{\end{center}} 
\def\ergs{\rm erg\,s^{-1}} 
 
\def\farcs{\hbox{$.\!\!^{\prime\prime}$}} 
\newcommand {\rc}{\rm}
\input psfigiallo.sty 
\begin{document} 
 
%\thesaurus{06 % A&A Section 6: Form. struct. and evolut. of stars
%              (08.09.2;  % stars: individual,
%               08.02.3;  % binaries: general,
%               08.16.6;  % pulsars: general
%               08.05.2); % Stars: emission-line, Be  
%               13.25.5;  % X-ray: stars,             
%               }

\title{The identification of the optical/IR counterpart of the  
29.5-s transient X--ray pulsar \src\thanks{Partially based on observations carried out at ESO, La Silla, Chile (65.H--0005)}}

\author{GianLuca Israel$^{1,}$\thanks{Affiliated to I.C.R.A.},  
Ignacio Negueruela$^{2,3}$, Sergio Campana$^{4,\star\star}$, Stefano Covino$^4$,  
Andrea Di Paola$^1$, Diana H. Maxwell$^5$,  
Andrew J. Norton$^5$,  
Roberto Speziali$^1$, Francesco Verrecchia$^6$ and Luigi Stella$^{1,\star\star}$}  

\institute{ 
Osservatorio Astronomico di Roma, Via Frascati 33,  
I--00040 Monteporzio Catone, Italy 
\and 
Observatoire Astronomique de Strasbourg,  
rue de l'Universit\'e 11, F67000 Strasbourg , France 
\and 
SAX Science Data Center, ASI c/o Telespazio, via Corcolle 19, I--00131, 
Roma, Italy  
\and
Osservatorio Astronomico di Brera, Via E. Bianchi 46,  
I--23807 Merate, Italy 
\and 
Dept. of Physics \& Astronomy, The Open University, Walton Hall,  
Milton Keynes MK7 6AA, U.K. 
\and 
Dipartimento di Fisica, Universit\`a degli Studi ``La Sapienza'', 
Piazza A. Moro 5, I--00185 Roma, Italy 
} 
 
\date{Submitted December 21, 2000; accepted March 21, 2001} 
\offprints{Gianluca.Israel@oar.mporzio.astro.it} 
\authorrunning{Israel et al.} 
\titlerunning{Identification of the optical/IR counterpart of \src} 
 
\abstract{ 
We report on the identification of the optical/IR counterpart  
of the 29.5-s transient X--ray pulsar \src.  
We re--analysed an archival \R\ HRI observation of \src\, obtaining a new  
refined position. The optical and IR follow--up  
observations carried out for the new error circle allowed us to find  
a relatively faint ($V$=20.9) and variable early type reddened star 
($V-R$=2.1). The  
optical spectra show the Balmer and Paschen series in emission, while the  
IR observations confirm the presence of a flux excess ($H$=13.2,  
$J-H$=0.54), suggesting that the star is surrounded by a  
circumstellar envelope.  
Spectroscopic and photometric data together indicate a B0--B2\,IV--Ve  
spectral--type star located at a distance of $\geq$\,10\,kpc confirming   
the Be--star/X--ray binary nature of \src. 
\keywords{stars: individual:  -- \src; \srchri\ -- 
                binaries: general --  
                stars: pulsars: general -- 
                stars: emission-line, Be -- 
                X-rays: stars  
                } }
\maketitle
 
\section{Introduction} 
Be/X--ray binary systems (BeXBs) represent the majority of the known   
High Mass X--ray Binaries (HMXBs) hosting an accreting rotating  
magnetic neutron star (White et al. 1995; Coe 2000).  
Based on the displayed X--ray features, BeXBs can be divided into at least   
three subclasses: (i) bright transients which display giant X--ray outbursts  
up to L$_{\rm X}=10^{38}$\,$\ergs$ (Type II; Stella et al. 1986) 
unrelated to the  
orbital phase, with high spin--up rates, (ii) transients  which  
display periodic outbursts of relatively high luminosity  
(L$_{\rm X}\simeq 10^{36}$\,--\,$10^{37}$\,$\ergs$; Type I) 
generally occurring  close to the periastron passage of the neutron 
star, and (iii)  sources displaying no outbursts, but comparatively 
moderate variations (up to a factor of $\sim$ 10--100) and 
low--luminosity ($\leq10^{36}$\,$\ergs$) pulsed persistent emission 
(Negueruela 1998). 
B--emission (Be) spectral--type stars are characterized by high rotational  
velocities (up to 70\% of their break--up velocity), and by episodes 
of equatorial mass loss which may produce a ring of gas orbiting  
around the star at irregular time intervals. At optical wavelengths, 
Be stars are difficult to  classify due to the presence of the 
circumstellar envelope responsible for the emission lines.

\begin{table*}[bht] 
\begin{center} 
\caption{Optical/IR observations carried out for the field of \src.} 
\begin{tabular}{lcllcl} 
\hline  
Telescope \& Instrument& Date & Exp.   & Seeing & Range  & Comments\\ 
                       &       &  (s) & (\arcsec) &(Band/\AA) &  \\  
\hline  
1.0\,m JKT \&  TEK4    & 1997 July 14&60     &   1\farcs0   &$V$&  \\ 
~~------    &           ''&60 & ''&$R$\\ 
~~------    &           ''&20 & ''&$I$\\ 
1.0\,m JKT \&  SITe 2  & 1999 July 24&500     &  1\farcs1   &$V$&  \\ 
~~------    &           ''&60 & ''&$R$\\ 
~~------    &           ''&20 & ''&$I$\\ 
3.5\,m NTT  \&  SUSI2  & 2000 May 5&300    & 2\farcs0 &$VRI$& \\ 
~~------                     &           ''&600    & ~ ''      &$B$& \\ 
1.1\,m AZT--24 \& SWIRCAM & 2000 June 3&30 & 2\farcs1 & $JH$ & \\ 
1.5\,m Cassini \&BFOSC &2000 June 30 &5400 & 1\farcs8 &4000--9000& Res. 15\AA\\ 
1.5\,m Danish \& DFOSC &2000 July 27 &10800& 0\farcs9 &4000--9000& Res. 15\AA \\ 
~~------           &2000 July 28 &3600 & 1\farcs0 &4000--9000&    '' \\ 
4.2\,m WHT \& ISIS   &2000 August 20 &3600 & 1\farcs0     &4000--10000& Res. 3.3--6\AA \\ 
~~------           &2000 August 21 &1800 & cloudy     &'' & ''\\ 
  \hline  
\end{tabular}   
\end{center} 
\end{table*}  
 
The optical counterparts of the Galactic X--ray transient 
sources suffer from high absorption columns and reddening, which 
hampers the possibility of detecting and classifying them. 
So far only about $\sim$20 optical counterparts of BeXBs have been 
discovered out of the $\sim$100 known X--ray pulsars (Negueruela 1998).  
A way around this problem is to observe the field of these X--ray sources  
in the IR; in fact, the bulk of the emission from matter around the  
Be peaks at IR wavelengths. 
Therefore, IR together with optical photometry make it easy  
to distinguish highly reddened but intrisically bright objects 
(i.e. late O -- early B stars) from nearby (low/no  
reddening) bright late type stars (such as K or M stars).  
Over the last three years our group has carried out (mainly by using ESO 
telescopes) an intensive observational campaign aimed at unveiling the 
optical/IR counterparts of an X--ray source sample made up by either 
recently discovered X--ray pulsators or by well-known transient X--ray 
pulsars with unknown companion. The project, so far, has led us to the  
unambiguous {\rc discovery} of the optical counterparts to  
$\sim$10 objects  (e.g. GS\,0834$-$43, Israel et al. 2000a; XTE\,J1946+274, 
Verrecchia et al. 2001; AX\,J1820.5$-$1434, Israel et al. 2000b). In 
some cases $V$ magnitudes in the 20--22 range and $V-R>2$ have been 
measured in our sample of identified optical/IR counterparts.   
 
The transient X--ray pulsar \src\ was discovered on 1988 April 
3 by {\em GINGA} intruments during a galactic plane scan 
observation near the Scutum region  
(\cite{Makino88a}; 1988b) and later confirmed through pointed  
observations carried out on 1998 April 19 and 
20. Coherent pulsations at a period of $P=29.5056\pm0.0002$\,s  
(pulsed fraction $\sim$7\%)  
were discovered in its X--ray flux, and a distance of at least 
5\,kpc  was proposed (\cite{Koyama90a}; 
\cite{Koyama90b}). The source was detected again during an outburst on 
1997 March 3 by   
the Burst and Transient Source Experiment ({\em BATSE}) on board CGRO  
(\cite{Wilson97}). A new measurement of the pulse period,  
$P=29.5631\pm0.0003$\,s, allowed a period derivative to be inferred of 
$\dot{P} = (-3.65\pm0.11)\times\,10^{-8}$\,s\,s$^{-1}$.  
 
A pointed observation carried out on 1997 March 5 by the {\em RXTE} 
Proportional Counter Array (PCA) found the source at a  
flux level of 60 mCrab (2--60\,keV; \cite{Takeshima97}).  
On 1997 April 4 the {\em BeppoSAX} Narrow Field Instruments 
(NFIs) performed a 
pointed observation of \src\ (\cite{Piraino98}; 2000).  
By exploiting the spatial resolution of the {\em BeppoSAX} 
imaging  instruments, the 90\% uncertainty for the position of \src\ 
was constrained to be within a 30\arcsec\ radius circular error region  
centered 
on {\rc ${\rm \rc RA = 18^{h} 45^{m} 34^{s} \pm 2^{s}}$, 
${\rm \rc Dec = 0^{\circ} 52\farcm5 \pm 0\farcm5}$} (equinox 2000; 
\cite{Santangelo97}). On the same day, the source 
was also observed with the 
ROSAT High Resolution Imager (HRI) which inferred a position 
of {\rc ${\rm \rc RA = 18^{h} 45^{m} 36\fs9 \pm 0.^{s}6}$, 
${\rm \rc Dec = 0^{\circ} 51' 45'' \pm 10''}$ (90\% confidence level 
uncertainty, 
equinox 2000; \cite{Dennerl97})}. 
 
\section{X--ray position}  
 
The positions for \src\ inferred from the {\em BeppoSAX} 
and the {\em ROSAT} observations which were carried out within a few 
hours of each other, are only marginally consistent {\rc ($\sim$63\arcsec\ 
distance each other)}. In order to {\rc reconcile} this difference  
we re--analysed both observations. The {\em ROSAT} HRI position was 
extracted from the Brera Multiscale Wavelet (BMW) catalog obtained by 
means of a wavelet--based detection technique developed to analyse 
high energy astronomical images (Lazzati et al. 1999; Campana et 
al. 1999). Only one source, namely 1BMW\,184536.8+005148, was found 
within the HRI field of  
view (475\,s of effective exposure time). The position of the source,   
${\rm RA = 18^{h} 45^{m} 36\fs8}$, ${\rm Dec = 0^{\circ} 51' 47.5''}$  
(90 \% confidence error radius $9''$, equinox 2000), is nearly coincident  
to that inferred by Dennerl \& Greiner (1997).  
 
%FIGURE 1 
%\vspace{1cm} 
\begin{figure*}[htb] 
\centerline{\psfig{figure=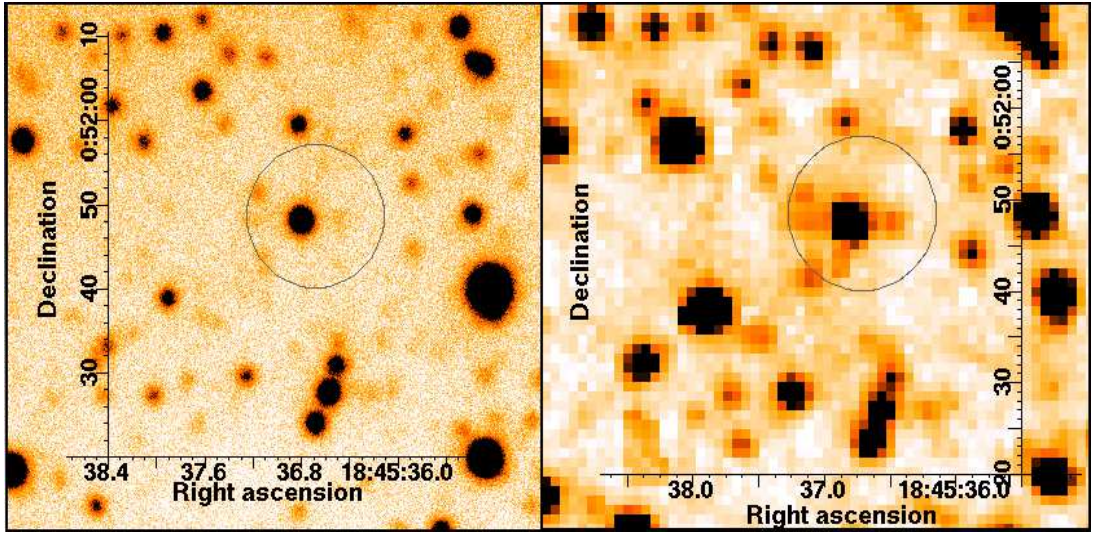,width=14.4cm,height=7cm} } 
\caption{Optical $R$--filter (left panel; NTT) and IR $H$--filter 
(right panel; AZT--24) images of the field of \src\ together with the new 
X--ray position uncertainty region  
(9\arcsec\ radius) derived from the 1997 April 4 \R\ HRI observation.  
North is top, east is left. The relatively bright star within the circle  
is the proposed optical counterpart of \src.} 
\end{figure*} 
 
{\rc Although also for the {\em BeppoSAX} observation we obtained 
nearly the same position reported by Santangelo et al. (1997), we note  
that in this case the  
uncertainty radius at 90\% cannot be set to 30\arcsec\ due to the relatively  
poor aspect spacecraft solution during this observation (the {\tt STR\_CONF}  
star traker flag in the house--keeping data $\leq$ 3) and, therefore, a 
more realistic value in the 50\arcsec--70\arcsec\ range must be considered.   
In the following we will refer to the ROSAT HRI position obtained from the 
BMW catalog and by Dennerl \& Greiner (1997) as the correct one.}

\section{Optical/IR observations} 

We observed the field including the ROSAT HRI error circle several times 
during 1997--2000 both in the optical and IR band (see Table\,1). 
The field was first imaged at optical 
wavelengths ($VRI$ filters) on 1997 July 14 with the 1.0-m 
Jacobus Kapteyn Telescope (JKT; La Palma) equipped with the TEK4 CCD 
(5\farcm6\,$\times$\,5\farcm6 field of view and 0\farcs33/pixel 
resolution). A further set of $VRI$ images was obtained at the JKT on 
1999 July 24 with the SITe2 CCD  
(10\arcmin$\times$\,10\arcmin\ field of view and 0\farcs33/pixel resolution). 
In both cases Starlink's {\it CCDPACK} (Draper 2000) was used to 
perform the bias subtraction and  flat--field correction, while 
GAIA (Draper \& Gray 2000) was employed for the photometry. 

Optical images in the $B$ (600\,s) and $VRI$ (300\,s each) filters were  
taken on 2000 May 5 with the ESO 3.5-m New Technology Telescope 
(NTT) at La Silla equipped with the Superb  Seeing Imager  -- 2 
(SUSI2; 5\farcm5\,$\times$\,5\farcm5 field of view and 
0\farcs16/pixel resolution). 
The data were reduced using standard {\sc ESO--MIDAS} and {\sc IRAF} 
procedures for bias subtraction and flat--field correction. Photometry 
for each stellar object in the image was derived with the DAOPHOT\,II 
program (Stetson 1987). 
Astrometry of the images was performed by using the USNO star catalog  
which provides an absolute positional accuracy of $\sim$0\farcs5 (Monet 1998). 
The \src\ field was further observed in the $J$ (30\,s) and $H$ (30\,s)  
filters on 2000 June 3   
from the 1.1-m AZT--24 telescope at Campo Imperatore (Italy) equipped with  
the Supernova Watchdogging IR Camera (SWIRCAM; 
4\farcm4\,$\times$\,4\farcm4 field of view and 1\farcs04/pixel 
resolution). {\rc Data analysis} procedures similar to those described above 
were applied. Within the X--ray positional uncertainty circle we  
detected only one highly reddened ($V-R=2.09$) and relatively bright 
($H=13.2$) IR object (see Fig.\,1) as expected for a distant and 
instrinsically blue early--type star. Moreover the star is the only 
variable object in the error circle
Table\,2 lists the results of the optical and IR photometry for the proposed  
optical counterpart of \src. 
 
\begin{table*}[htb]  
\begin{center}  
\caption{Optical and IR results for the proposed counterpart.}  
\begin{tabular}{lllllclllc}  
\hline  
Date & $B$ & $V$ & $R$ & $I$ & $V-R$ && $J$ & $H$ & $J-H$ \\  
\hline  
July 1997 & -- & 20.6$\pm$0.5 & 18.6$\pm$0.2 & 16.5$\pm$0.2 & 2.0& & -- & -- & --  \\ 
July 1999 & -- & 20.2$\pm$0.3 & 18.5$\pm$0.3 & 17.1$\pm$0.3 & 1.7& & -- & -- & -- \\ 
 May 2000 & $>$24.1 & 20.89$\pm$0.05  & 18.80$\pm$0.05  & 16.79$\pm$0.05  & 2.09 && -- & -- & -- \\ 
June 2000 & -- & -- & -- & -- & -- && 13.75$\pm$0.05 & 13.21$\pm$0.05 & 0.54 \\ 
\hline \\ 
\multicolumn{2}{l}{R.A.(J2000) = }& 
\multicolumn{8}{l}{~~\,18$^{\rm h}$ 45$^{\rm m}$ 36\fs8}\\ 
\multicolumn{2}{l}{Dec.(J2000) = } &  
\multicolumn{8}{l}{+00$^{\circ}$ 51\arcmin\ ~48\farcs3}\\ 
\multicolumn{2}{l}{Spectral type:} &  
\multicolumn{8}{l}{~~\,B0--B2\,IV--V\,e}\\ 
\multicolumn{2}{l}{$E_{B-V}$:} &  
\multicolumn{8}{l}{~~2.3$\,\rightarrow\,$2.9}\\ 
\multicolumn{2}{l}{Distance:} &  
\multicolumn{8}{l}{~~$\geq$10\,kpc}\\  
&&&&&&&\\ 
\hline 
\end{tabular}  
\end{center}  
Note --- $B$ magnitude upper limit is at  
90\% confidence level. Position uncertainty is 0\farcs5.\\  
\end{table*}  
%FIGURE 2 
%\vspace{1cm} 
\begin{figure*}[htb] 
\centerline{\psfig{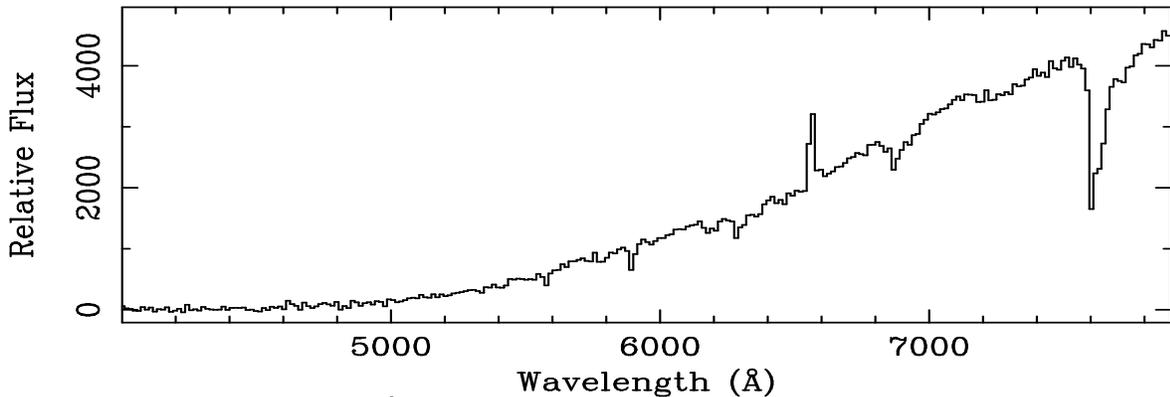} } 
\caption{The 14\,400-s low resolution (15\AA) spectrum of \src\ obtained on 2000 July  
27--28 from the 1.5\,m Danish telescope at La Silla (Chile).} 
\end{figure*} 
 
Spectroscopic observations of the candidate optical counterpart were 
obtained on 2000 June 30 with the 1.52\,m ``Cassini" telescope (Loiano 
Observatory) equipped with the Bologna Faint Objects 
Spectrometer and Camera (BFOSC; Bregoli et al. \cite{BFMFO87}; 
Merighi et al. \cite{MMCMBO94}; 12\farcm2\,$\times$\,12\farcm2 field 
of view and 0\farcs58/pixel resolution). We performed low--resolution (15\AA) 
spectroscopy for a total exposure time of 5\,400\,s (3$\times$\,1\,800\,s). 
After applying standard corrections, cosmic rays were removed 
from each spectrum and the sky--subtracted stellar spectra were obtained, 
corrected for atmospheric extinction. The signal to noise (S/N) ratio was 
extremely low and 
object counts were only detected above 5500\AA. A strong emission line 
(equivalent width, EW, of --14$\pm$2\AA\ and Full Width Half Maximum, 
FWHM, consistent with the instrumental resolution) was 
detected, its wavelength corresponding to that of H$\alpha$. This 
result strongly suggested that the optical source was associated 
with \src.    
In order to achieve better statistics we further observed the candidate 
optical counterpart to \src\ on 2000 July 26 and 27 from 
the ESO 1.5-m Danish telescope (La Silla, Chile) equipped with the 
Danish Faint  Object Spectrometer Camera (DFOSC; 
15\arcmin\,$\times$\,15\arcmin\ field of view and 0\farcs4/pixel resolution). 
Five low--resolution (15\AA) spectra were obtained for a total effective 
exposure time of 14\,400\,s. Reduction procedures similar to those 
described above were applied to each spectrum. The resulting summed 
spectrum (flux uncalibrated) is shown in Fig.\,2. Also in this case 
the H$\alpha$ emission line was evident (EW of --14$\pm$1\,\AA), 
while no other emission lines were detected. 

Finally on 2000 August 20 and 21 two medium--resolution spectra 
(total on--source integrations of 3\,600\,s and 1\,800\,s respectively) were 
taken with the Intermediate Dispersion Spectroscopic and Imaging System 
(ISIS) mounted on the 4.2\,m William Herschel Telescope (WHT), located at the 
Observatorio del Roque de los Muchachos, (La Palma, Spain). The blue arm was 
equipped with the R158B grating and the EEV\#10 CCD, which gives a nominal 
dispersion of $\sim$1.6\AA/pixel.
The resolution at $\sim$6000\AA, estimated from the FWHM of arc 
lines, is $\sim$6\AA. The red arm was equipped with the R316R grating and 
the Tek4 CCD, which gives a nominal dispersion of $\sim$1.5\AA/pixel 
(the resolution is $\sim$3.3\AA\ at $\sim$8000\AA). 

The $I$--band spectrum of the optical counterpart to \src\  
displays the wealth of emission lines typical of early--type Be stars, 
which is generally seen in Be/X-ray binary counterparts (Negueruela 
\& Torrej\'{o}n, 2001; see Fig.\,3). The Paschen series appears 
strongly in emission from Pa11 down to Pa22 (though the last members 
of the series are difficult to measure due to increasing atmospheric 
contamination). Parameters of several emission lines are displayed in 
Table\,\ref{tab:lines}. Pa18 is clearly double--peaked, indicating blend 
with \ion{O}{i}~$\lambda$8446\AA. The larger FWHMs and EWs of Pa15 
and Pa16 indicate that they are also blended with the 
\ion{Ca}{ii}~$\lambda\lambda$~8498, 8542~\AA\ lines, though there is 
nothing in the parameters of Pa13 to make us suspect its blending with 
the third member of the \ion{Ca}{ii} triplet, at $\lambda$8662\AA. 
The \ion{N}{i} band at $\lambda\lambda$~8680\,--\,8686~\AA, sometimes 
seen in emission in Be stars (Andrillat et al. 1988) could be on the 
red wing of Pa13, but the {\rc S/N ratio} is not enough to ascertain it. 

%FIGURE 3 
%\vspace{1cm} 
\begin{figure*}[htb] 
\centerline{\psfig{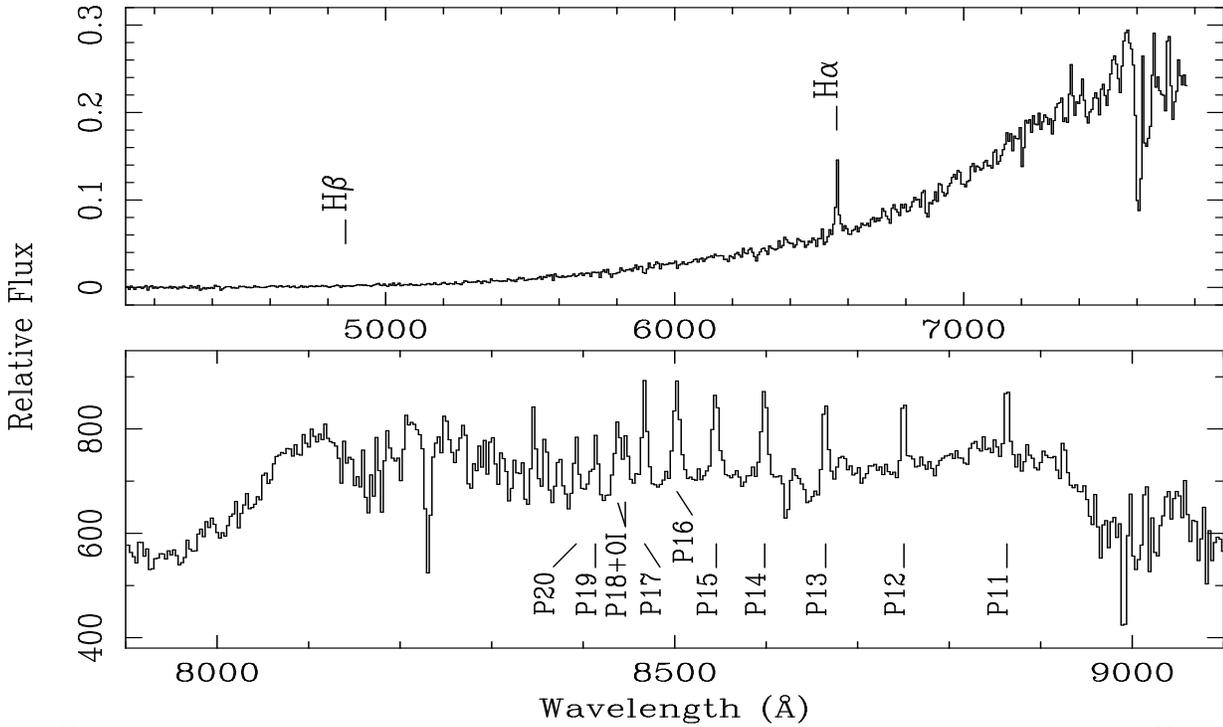} } 
\caption{The medium resolution spectra of \src\ obtained on 2000 August 20  
and 21 at the WHT at La Palma (Spain).} 
\end{figure*} 
 
Apart from H$\alpha$, no other lines are observed in the blue part of 
the optical spectrum. The flux shortwards of $\lambda6000$\AA\ is too weak for 
any line to be detected. The 
\ion{O}{i}~$\lambda\lambda$~7772\,--\,7775~\AA\ band,  
which is sometimes seen in emission in Be stars, falls in the region affected  
by the dichroic. The lack of any spectral absorption features prevents us  
from an accurate spectral classification. However, the presence of 
strong emission in the Paschen series identifies the star as earlier 
than B2 (Andrillat et al. 1988), i.e., in the spectral range occupied 
by known Galactic Be/X--ray binary counterparts (Negueruela 1998).  
 
Due to the difficulty in determining the local continuum, the values 
of EWs have large uncertainties, typically $\sim 15$\%, but as high as 
30\% for Pa11 (uncertainties have been estimated by considering 
different continuum levels). The narrowness of all the unblended lines 
as well as their shape, clearly shows that this Be star is close to 
pole--on. Be stars with larger inclinations to the line--of--sight 
generally display double--peaked Paschen emission lines with $\Delta 
v_{{\rm peak}} > 200\:{\rm km}\,{\rm s}^{-1}$, i.e., clearly separable 
at this resolution.

\begin{table}[htb] 
 \caption{Measured parameters of emission lines detected in the  
WHT spectra of the proposed optical counterpart to \src.  
The FWHMs have been corrected for instrumental effects.  
See text for details.} 
\begin{center} 
\begin{tabular}{lcc} 
\hline 
Line & EW & FWHM \\ 
     & (\AA) & (km s$^{-1}$) \\ 
\hline 
Pa 11 & $-$1.0 & 135\\ 
Pa 12 & $-$1.3 & 115\\ 
Pa 13 & $-$1.2 & 140\\ 
Pa 14 & $-$1.8 & 220\\ 
Pa 15 & $-$2.0 & 300\\ 
Pa 16 & $-$2.2 & 285\\ 
Pa 17 & $-$2.1 & 160\\ 
Pa 18 + \ion{O}{i}~$\lambda$8446\AA& $-$2.5 & blend\\ 
Pa 19 & $-$1.3 & $-$\\ 
Pa 20 & $-$1.2 & 170\\ 
H$\alpha$ & $-$18 & 280\\ 
\hline 
\end{tabular} 
\end{center} 
 \label{tab:lines} 
\end{table}

\section{Discussion and Conclusion} 
The X--ray, optical and IR observations of the field of \src\ presented  
here led to the identification of the optical counterpart of this 29.5\,s  
transient X--ray pulsar discovered in 1988. The measurement of the distance  
to \src\ based on the optical data is hampered by the uncertainties in the  
spectral classification.  
 
However some information can be inferred based on our optical and IR 
photometric  
measurements. The intrinsic $(V-R)$ color for \src\ is $\sim -0.1 
\rightarrow -0.15$ (assuming a main sequence or sub--giant star with 
spectral type in  the B0--B2 range).  
Since the observed $(V-R)$ is $\sim1.7\rightarrow2.1$ the reddening should  
amount to $E_{V-R}\sim1.8\rightarrow2.3$, and assuming a standard reddening  
law (Fitzpatrick 1999) this converts to $A_{R}\sim5.6\rightarrow7$,  
$A_{V}\sim7\rightarrow9$ and $E_{B-V}\sim2.3\rightarrow2.9$ (regardless  
of whether the reddening medium is uniformly distributed along the line of  
sight or is intrinsic to the source). However from the X--ray spectral fits  
(Piraino et al. 2000) a $N_H$ of 2--3\,$\times10^{22}$\,cm$^{-2}$ was 
derived, corresponding to an  $E_{B-V}\sim3.2\rightarrow5$ (Bohlin et 
al. 1978), which is not consistent with the above results. Considering 
the total Galaxy $N_H$ column in the direction of \src, 
1--2$\times10^{22}$ cm$^{-2}$, a   
value of $E_{B-V}\sim2.7\rightarrow3.2$ was inferred. This result  
suggests that at least part of the inferred X--ray $N_H$ is 
local to the system and obscures the neutron star during outbursts and 
that the binary system is quite far from us. A good agreement with the 
$V$, $R$ and $I$ measurements is obtained for a B1IV--V star at a distance 
larger than 10\,kpc (note that in the direction to \src\ the Galaxy 
edge is located at $\sim15$\,kpc) and an  
$E_{B-V}\sim\,2.8$. However, according to Hayakawa  et al. (1977), the  
interstellar extinction to the direction of \src\ is somewhat larger 
than the average Galactic plane value, which might imply a 
smaller distance (in the 5--8\,kpc range) to the source. We can 
reasonably discard the possibility  
of a luminosity class III which would imply an $E_{B-V}\sim3.2$ 
and an IR--deficiency. For a reference B1V--IV star ($M_{V}\sim-3.5$) 
at a distance of 10--15\,kpc and based on the IR photometry we infer 
an excess $\geq1.5$ and $\geq1.2$ magnitudes in $J$ and $H$ filters, 
respectively, suggesting the presence of a circumstellar envelope.  
{\rc We also note that the proposed counterpart is the only object 
displaying flux variability in the field of \src\ ($\sim2\sigma$ 
confidence level in the $V$ and $I$ band) as expected for Be/X--ray binaries, 
further suggesting the correctness of the identification. }
 
Based on both photometric and spectroscopic findings we conclude that the  
proposed optical counterpart of \src\ is most likely a B0--2\,V--IVe variable  
star at a distance larger than 10\,kpc. A more accurate distance and spectral 
classification would require more detailed optical spectroscopic observations 
in the blue end of the spectrum with a larger telescope. 
 
For  a distance of 10--15\,kpc and a 1--10\,keV flux of  
$\sim$\,6\,$\times$\,10$^{-10}$\,erg~s$^{-1}$~cm$^{-2}$ as  
measured by BeppoSAX during the 1997 April outburst (Piraino et al. 2000)  
we obtain an X--ray luminosity of 
L$_{\rm X}$\,(1--10\,keV)\,$\simeq$\,5--20\,$\times$\,10$^{36}$\,erg~s$^{-1}$.  
Such a luminosity is a typical value shown by X--ray pulsars in binary 
systems during Type I outbursts (Stella et al. 1986; Negueruela 1998)   
occurring close to the time of periastron passage and with a periodic  
recurrence at the orbital period of the system.

\begin{acknowledgements}  
This work is supported through CNAA, ASI and Ministero dell'Universit\`a e 
Ricerca  Scientifica e Tecnologica (MURST) grants. IN and GLI thank  
the service observing programme at the WHT in La Palma. AJN and DHM 
thank the service observing programme at the JKT in La Palma. During 
part of this work, IN was supported by an ESA external fellowship. 
The WHT and JKT are operated on the 
island of La Palma by the Royal Greenwich Observatory in the Spanish 
Observatorio del Roque de Los Muchachos of the Instituto de
Astrof\'{\i}sica de Canarias.
\end{acknowledgements} 
          
\newpage


\begin{thebibliography}{} 
 
\bibitem{ } Andrillat, Y., Jaschek, M., Jaschek, C. 1988, A\&AS, 72, 129 
 
\bibitem{ } Bohlin, R.C., Savage, B.D., \& Drake, J.F. 1978, ApJ, 224, 132 
 
\bibitem[1987]{BFMFO87} Bregoli, G., Federici, L., Merighi, R., et al. 1987, 
in: ESO--OHP  Workshop on the Optimization of the Use of CCD Detectors in 
Astronomy,  Saint--Michel--l'Observatoire, France, June 17--19, 1986, 
Proceedings (A88--13301  03--89). Garching, ESO, Germany, p. 177 
 
\bibitem{ } Campana, S., Lazzati, D., Panzera, M.R., et al. 1999, 
ApJ, 524, 423 
 
\bibitem{ } Coe, M.J. 2000, Proceedings of the  
175$^{th}$ IAU Colloq., Eds. M.A. Smith, H.F. Henrichs \& J. Fabregat, A.S.P. Conf.  
Ser. Vol. 214, p.\,656
 
\bibitem[Dennerl \& Greiner 1997]{Dennerl97}Dennerl, K., Greiner, J. 1997, 
IAU  Circ., 6645 
 
\bibitem{ } Draper, P.W., Taylor, M., Allan, A. 2000, Starlink User 
Note 139.12,  R.A.L.  
 
\bibitem{ } Draper, P.W., Gray, N. 2000, Starlink User Note 214.7, R.A.L. 
 
\bibitem{ } Hayakawa, S., Ito, K, Matsumoto, T., et al. 1977, A\&A, 56, 325 

\bibitem{ } Fitzpatrick, E.L. 1999, PASP, 111, 63
 
\bibitem{ } Israel, G.L., Covino, S., Campana, S., et al. 2000a, 
MNRAS, 314, 87 
 
\bibitem{ } Israel, G.L., Covino, S., Polcaro, V.F., et al. 2000b, 
Proceedings of the  175$^{th}$ IAU Colloq., Eds. M.A. Smith, H.F. Henrichs 
\& J. Fabregat, A.S.P. Conf.  Ser. Vol. 214, p.\,739  
 
\bibitem[Koyama et al. 1990a]{Koyama90a}Koyama, K., Kawada, M., Kunieda, H., 
et al. 1990a, Nature, 343, 146 
 
\bibitem[1990b]{Koyama90b}Koyama, K., Kawada, M., Takeuchi, Y., et al. 1990b, 
ApJ,  356, L47 
 
\bibitem{ } Lazzati, D., Campana, S., Rosati, P., et al. 1999, ApJ, 524, 414 
 
\bibitem[Makino et al. 1987]{Makino87}Makino, F., and the ASTRO--C Team  
1987, Ap. Letters Comm., 25, 223 
 
\bibitem[Makino et al. 1988a]{Makino88a}Makino, F., and {\em GINGA} 
Team 1988a, IAU Circ., 4583 
 
\bibitem[Makino et al. 1988b]{Makino88b}Makino, F., and {\em GINGA} Team  
1988b, IAU Circ., 4587 
 
\bibitem[1994]{MMCMBO94} Merighi, R., Mignoli, M., Ciattaglia, C., et al.  
        1994, Bologna Technical Reports 09,-1994,-05 
 
\bibitem{ } Monet, D.G. 1998, AAS Meeting \#193, \#120.03 
 
\bibitem{ } Negueruela, I. 1998, A\&A, 338, 505 
 
\bibitem{ } Negueruela, I., Torrej\'{o}n, J.M. 2001, in prep.  
 
\bibitem[Piraino et al. 1998]{Piraino98}Piraino, S., Santangelo, A., 
Giarrusso, S., et al. 1998, Nucl. Phys. B (proc. Suppl.), 69/1--3, 220 
 
\bibitem[Piraino et al. 2000]{Piraino00}Piraino, S., Santangelo, A., 
Segreto, A., et al. 2000, A\&A, 357, 501 
 
\bibitem[Santangelo et al. 1997]{Santangelo97}Santangelo, A., Giarrusso, S., 
Del Sordo, S., et al. 1997, IAU Circ., 6659 
 
\bibitem{ } Stella, L., White, N.E., Rosner, R. 1986, ApJ, 208, 669 
 
\bibitem{ } Stetson, P. B. 1987, PASP, 99, 191 
 
\bibitem[Takeshima 1997]{Takeshima97}Takeshima, T. 1997, IAU Circ., 6595 
 
\bibitem{ }Verrecchia, F., Israel, G.L., Negueruela, I. 2001, in prep.  
 
\bibitem[Wilson et al. 1997]{Wilson97}Wilson, R.B., Harmon, B.A., Scott, 
D.M., et al. 1997, IAU Circ., 6586 

\bibitem{ } White, N. E., Nagase, F., Parmar, A. N. 1995 , in
{\it X--ray Binaries}, Eds. W.H.G. Lewin, J. van Paradijs \&
E.P.J. van den Heuvel (Cambridge University Press), p.1 
 
\end{thebibliography}
\end{document}